\documentclass[aps,prb,preprint,showpacs,eqsecnum]{revtex4-1}
\usepackage{graphicx}
\usepackage{bm}
\usepackage{amsmath,amssymb}
\begin{document}
\title{Induced magnetization and power loss for a periodically driven system of ferromagnetic nanoparticles with randomly oriented easy axes}
\author{S.~I.~Denisov}
\email{denisov@sumdu.edu.ua}
\author{T.~V.~Lyutyy}
\email{lyutyy@oeph.sumdu.edu.ua}
\author{B.~O.~Pedchenko}
\author{O.~M.~Hryshko}
\affiliation{Sumy State University, 2 Rimsky-Korsakov Street, UA-40007 Sumy, Ukraine}


\begin{abstract}
We study the effect of an elliptically polarized magnetic field on a system of non-interacting, single-domain ferromagnetic nanoparticles characterized by a uniform distribution of easy axis directions. Our main goal is to determine the average magnetization of this system and the power loss in it. In order to calculate these quantities analytically, we develop a general perturbation theory for the Landau-Lifshitz-Gilbert (LLG) equation and find its steady-state solution for small magnetic field amplitudes. On this basis, we derive the  second-order expressions for the average magnetization and power loss, investigate their dependence on the magnetic field frequency, and analyze the role of subharmonic resonances resulting from the nonlinear nature of the LLG equation. For arbitrary amplitudes, the frequency dependence of these quantities is obtained from the numerical solution of this equation. The impact of transitions between different regimes of regular and chaotic dynamics of magnetization, which can be induced in nanoparticles by changing the magnetic field frequency, is examined in detail.
\end{abstract}
\maketitle

\section{INTRODUCTION}
\label{Intr}

The study of the magnetization dynamics in single-domain ferromagnetic particles (nanoparticles) is of large importance for both fundamental research and practical applications. The dynamical effects related to a sudden change in the magnetization behavior, which occurs as a control parameter is varied, are of the greatest interest. One of these effects is the switching (or reversal) of the nanoparticle magnetization. Depending on the application, the switching process has to be properly optimized. In particular, to reduce the switching time and switching magnetic field in magnetic recording devices, the so-called precessional switching\cite{BFHS, GBHB, KaRa, SMB, SuWa1} and microwave-assisted switching\cite{TWM, SuWa2, WoBa, ZZT, BMSA, PGBT} have recently been proposed. The magnetic resonances in nanoparticles and transitions between different dynamical states of the magnetization can also play an important role in hyperthermia,\cite{PCJD, ISHK, LFPR, LDHM} because a strong change of nanoparticle heating is expected to occur in the vicinity of these resonances and transitions. A remarkable feature of the deterministic dynamics of magnetization in nanoparticles driven by periodic magnetic fields is that it can be chaotic. \cite{APC, BPSV, LBGP, SQBA} From a theoretical point of view, the transitions between the regular and chaotic regimes of the magnetization dynamics and routes to chaos have a special interest.

The nonlinear magnetization dynamics in ferromagnetic nanoparticles driven by a circularly-polarized magnetic field is well studied for the particular case of uniaxial nanoparticles, whose easy axes are perpendicular to the polarization plane\cite{BSM, BMS1} (see also Ref.~[\onlinecite{BMS}] and references therein). Using the deterministic Landau-Lifshitz-Gilbert (LLG) equation,\cite{LaLi, Gilb} these authors have shown that the magnetization dynamics is always regular and, in the steady state, only periodic and quasiperiodic dynamical regimes exist. In addition, the stability conditions for periodic regimes and induced magnetization are derived in,\cite{DLHT} the phase diagram of possible regimes in the plane `amplitude--frequency' of the driving magnetic field is obtained in,\cite{DLBH} the power loss for periodic regimes is calculated in,\cite{NaRa, RCSS} and the influence of transitions between different dynamical regimes on the power loss is studied in.\cite{LDPB} Some thermal effects in such systems, including thermal enhancement of the induced magnetization and resonant suppression of thermal stability of these regimes, are investigated in.\cite{DLH, DPL}

Due to the symmetry of the model, many of the above results were obtained analytically. At the same time, a number of important features of the magnetization dynamics (e.g., some higher-order resonances and chaotic dynamics) are symmetrically forbidden in this model. Therefore, in this paper we consider a more general case when the driving field is elliptically polarized and the nanoparticle easy axis has a random direction. Here, our interest is focused on understanding how the nonlinear resonances and transitions between different regimes of the magnetization dynamics affect the magnetic properties of nanoparticle systems.

The paper is structured as follows. In Sec.~\ref{Gen}, we describe the model, introduce the basic equations, and define the quantities of interest. A general perturbation theory for the LLG equation is developed in Sec.~\ref{Pert}. In the same section, we determine the steady-state solution of this equation in the first and second orders of the perturbation theory. Section \ref{Magn} is devoted to studying the average magnetization of the reference system induced by the elliptically polarized magnetic field. The dependence of the power loss on the magnetic field amplitude and frequency is studied in Sec.~\ref{Power}. Finally, our results are summarized and discussed in Sec.~\ref{Concl}.

\section{GENERAL FRAMEWORK AND BASIC EQUATIONS}
\label{Gen}

We consider a system of ferromagnetic nanoparticles driven by the magnetic field $\mathbf{H} = \mathbf{H}(t)$, which is elliptically polarized in the $xy$ plane, i.e.,
\begin{equation}
    \mathbf{H} = H\cos{(\omega t)} \mathbf{e}_{x}
    + \rho H \sin{(\omega t)} \mathbf{e}_{y}.
    \label{H}
\end{equation}
Here, $H$ and $\omega$ are, respectively, the amplitude and angular frequency of the magnetic field, $\mathbf{e}_{x}$, $\mathbf{e}_{y}$ and $\mathbf{e}_{z}$ are the unit vectors of the Cartesian coordinate system $xyz$, and $\rho$ is the dimensionless parameter. The sign of this parameter characterizes the direction of field rotation (at $\rho <0$ the magnetic field rotates in the clockwise direction and at $\rho >0$ in the counterclockwise direction), and its values $\rho =0$ and $|\rho| =1$ correspond to the linearly and circularly polarized magnetic fields.

The nanoparticles of the system are assumed to be non-interacting and single-domain (this is so-called Stoner-Wohlfarth particles\cite{StWo}), and distributed in an insulating matrix. The only difference between them is the direction of their anisotropy axes (easy axes) that, for each nanoparticle, is characterized by the unit vector
\begin{equation}
    \mathbf{e}_{a} = \sin{\theta_{a}} \cos{
    \varphi_{a}}\mathbf{e}_{x} +  \sin{\theta_{a}}
    \sin{\varphi_{a}}\mathbf{e}_{y} +  \cos{
    \theta_{a}} \mathbf{e}_{z},
    \label{e_a}
\end{equation}
where $\theta_{a}$ and $\varphi_{a}$ are the polar and azimuthal angles of $\mathbf{e}_{a}$. In the following analysis, we assume that the directions of this vector are random and uniformly distributed over the sphere. This means that the angles $\theta_{a}$ and $\varphi_{a}$ are also random and their joint probability density $P(\theta, \varphi)$ that $\theta_{a} = \theta$ and $\varphi_{a} = \varphi$ is given by
\begin{equation}
    P(\theta,\varphi) = \frac{1}{4\pi}\sin{\theta}.
    \label{P}
\end{equation}

The magnetic state of each nanoparticle in the system is described by the magnetization vector $\mathbf{M} = \mathbf{M}(t)$, whose dynamics is governed by the deterministic LLG equation\cite{LaLi, Gilb}
\begin{equation}
    \frac{d}{dt}\mathbf{M} = -\gamma \mathbf{M}
    \times \mathbf{H}_{\mathrm{eff}} + \frac
    {\alpha}{M} \mathbf{M} \times \frac{d}{dt}
    \mathbf{M}.
    \label{LLG}
\end{equation}
Here, $\gamma (>0)$ is the gyromagnetic ratio, $\alpha(>0)$ is the dimensionless damping parameter, $M = |\mathbf{M}| = \textrm{const}$, and the cross sign denotes the vector product. The effective magnetic field $\mathbf{H}_{\mathrm{eff}} = \mathbf{H}_{\mathrm{eff}}(t)$ acting on the magnetization is taken in the following form:
\begin{equation}
    \mathbf{H}_{\mathrm{eff}} = \frac{H_{a}}{M}
    {(\mathbf{M}\cdot \mathbf{e}_{a})}\mathbf{e}_{a}
    + \mathbf{H},
    \label{H_eff}
\end{equation}
where $H_{a}$ is the magnetic anisotropy field and the dot denotes the scalar product. Introducing the dimensionless magnetization, $\mathbf{m} = \mathbf{M}/M$ ($|\mathbf{m}| =1$), and the dimensionless effective magnetic field, $\mathbf{h}_{ \mathrm{eff}} = \mathbf{H}_{ \mathrm{eff}}/ H_{a}$, the LLG equation (\ref{LLG}) can be reduced to the form
\begin{equation}
    \ell\dot{\mathbf{m}} = -\mathbf{m}
    \times \mathbf{h}_{\mathrm{eff}} - \alpha
    \mathbf{m}\times (\mathbf{m} \times \mathbf{h}
    _{\mathrm{eff}}).
    \label{red_LLG}
\end{equation}
Here, $\ell = 1 + \alpha^{2}$, the overdot denotes differentiation with respect to the dimensionless time $\tau = \omega_{a}t$, $\omega_{a} = \gamma H_{a}$ is the characteristic angular frequency of the magnetization precession, and, according to (\ref{H}) and (\ref{H_eff}),
\begin{equation}
    \mathbf{h}_{\mathrm{eff}} = {(\mathbf{m}\cdot
    \mathbf{e}_{a})}\mathbf{e}_{a} + h\cos{(\Omega
    \tau)} \mathbf{e}_{x} + \rho h \sin{(\Omega
    \tau)} \mathbf{e}_{y}
    \label{h_eff}
\end{equation}
with $h=H/H_{a}$ and $\Omega = \omega/ \omega_{a}$ being, respectively, the dimensionless amplitude and dimensionless angular frequency of the driving magnetic field $\mathbf{H}$. Because the direction of $\mathbf{e}_{a}$ is random, the dynamics of $\mathbf{m}$ in different nanoparticles of the system is, in general, different. Therefore, the average magnetic properties of nanoparticles play a key role in describing the corresponding magnetic properties of such a system.

In this paper, we are interested in two characteristics of nanoparticles. The first is the average dimensionless magnetization, $\langle \overline{\mathbf{m}} \rangle$, induced by the elliptically polarized magnetic field. Here, the overbar denotes averaging over the dimensionless time interval $\mathcal {T}$,
\begin{equation}
    \overline{(\cdot)} = \frac{1}{\mathcal {T}}
    \int_{0}^{\mathcal {T}}d\tau (\cdot),
    \label{mean1}
\end{equation}
and the angular brackets denote averaging over all possible orientations of the unit vector $\mathbf{e}_{a}$,
\begin{equation}
    \langle(\cdot)\rangle = \int_{0}^{\pi}d \theta
    \int_{0}^{2\pi}d\varphi P(\theta,\varphi)(\cdot).
    \label{mean2}
\end{equation}
It should be noted that the choice of $\mathcal {T}$ depends on the dynamical regimes of $\mathbf{m}$. In particular, in the case of the steady-state dynamics the time interval $\mathcal {T}$ should be chosen as $\mathcal {T} = 2\pi /\Omega$. In contrast, if $\mathbf{m}$ exhibits chaotic dynamics, then the following condition should be satisfied: $\mathcal {T} \gg 2\pi /\Omega$. The second quantity of our interest is the reduced power loss defined as $q = \langle \overline{Q} \rangle /(\omega_{a} H_{a} MV)$, where $Q = V \mathbf{H}_{\mathrm{eff}} \cdot d\mathbf{M} /dt$ is the instantaneous power loss per nanoparticle of the volume $V$. Using the LLG equation (\ref{LLG}), this quantity can be written as follows:
\begin{equation}
    q = \alpha \langle \overline{\dot{
    \mathbf{m}}^{2}} \rangle.
    \label{def_q}
\end{equation}
If the nanoparticle system of the volume $\mathcal {V}$ contains $N$ nanoparticles, then the induced magnetization and power loss density for this system are expressed through the above introduced quantities $\langle \overline{ \mathbf{m}} \rangle$ and $q$ as $nMV\langle \overline{\mathbf{m}} \rangle$ and $n \omega_{a} H_{a} MVq$, respectively, where $n = N/ \mathcal {V}$ is the concentration of nanoparticles.

Let us now formulate the conditions under which this model is justified. First of all, we assume that the strength of the exchange interaction between spins is the largest energy scale in the model. In this case, the magnetization magnitude is approximately constant and the magnetization rotation can be described by the LLG equation (\ref{LLG}). Since the rotation is considered to be coherent, the nanoparticles should be single-domain. This implies that the nanoparticle diameter $d$ must be less than some critical value $d_{2}$ which, depending on the material, ranges from a few nanometers to several tens or even hundreds of nanometers (for example,  $d_{2} \simeq 4.7 \, \mathrm{nm}$ for $\mathrm{Ni_{08} Fe_{02}}$, $d_{2} \simeq 19 \, \mathrm{nm}$ for $\mathrm{Fe}$, and $d_{2} \simeq 480 \, \mathrm{nm}$ for $\mathrm{MnBi}$\cite{Guim}). In general, because of thermal fluctuations, the coherent rotation of magnetization in nanoparticles with $d < d_{2}$ is random. In the framework of the stochastic LLG equation, these fluctuations are usually accounted for by adding a Gaussian white noise to the effective magnetic field\cite{Brown} (for recent reviews see, e.g., Refs.~[\onlinecite{CKW}] and [\onlinecite{BMS}] and references therein). However, if the thermal energy $k_{B}T$ ($k_{B}$ is the Boltzmann constant, $T$ is the absolute temperature) is much less than the smallest energy scale in the system, $Vw$ ($w = M \min{(H_{a},H)}$ is the scale energy density), than thermal fluctuations can safely be neglected. This occurs at $d \gg d_{1} = (6k_{B}T/ \pi w)^{1/3}$, and thus the magnetization is homogeneous and its dynamics is approximately deterministic if the nanoparticle diameter satisfies the conditions $d_{1} \ll d < d_{2}$. The condition $d_{1} \ll d$ can also be used to evaluate the maximum temperature at which the deterministic approximation still holds (this is the case if the maximum temperature is less than the blocking temperature). Note that these conditions are not too restrictive and in some cases can be satisfied even at room temperatures.\cite{LDPB}

In addition, we use the approximation of non-inter\-acting nanoparticles, i.e., the average interparticle distance $R$ is assumed to be so large that the total magnetic field produced by the surrounding nanoparticles is negligibly small compared to the anisotropy and external magnetic fields. In the dipole approximation, this distance can roughly be estimated from the condition $(R/d)^{3} \gg M/\min{(H_{a},H)}$. At first sight, even with the above assumptions, the choice of the effective magnetic field in the form (\ref{H_eff}) is still not satisfactory. The reason is that $\mathbf{H}_{ \mathrm{eff}}$ does not contain the demagnetizing magnetic field, which always exists in ferromagnetic samples and, in general, can not be neglected. But in the case of single-domain particles of spherical shape the demagnetizing field equals $-(4\pi/3) \mathbf{M}$ and, since $\mathbf{M} \times \mathbf{M} =0$, this field does not influence the magnetization dynamics and can be ignored in $\mathbf{H}_{\mathrm{eff}}$. Note that our approach can also be applied to conducting nanoparticles. This is because the main effect of conductivity is the renormalization of the damping parameter $\alpha$.\cite{MLT, DLPB} Finally, according to the definition (\ref{mean2}), the quantities of our interest, $\langle \overline{\mathbf{m}} \rangle$ and $q$, depend on the probability density $P(\theta, \varphi)$ of easy axis directions. The choice of the uniform distribution for these directions is motivated by both physical (zero magnetization of non-driving systems) and mathematical (simple integration over the angles $\theta$ and $\varphi$) reasons. However, any other choice of $P(\theta, \varphi)$ is also possible; the only problem in this case is the analytical calculation of the integrals in (\ref{mean2}).

\section{PERTURBATION THEORY}
\label{Pert}

Assuming that $h\ll 1$, we represent the reduced magnetization $\mathbf{m}$ in the series form
\begin{equation}
    \mathbf{m} = \sum_{n=0}^{\infty}\mathbf{m}_{n},
    \label{ser_m}
\end{equation}
where $\mathbf{m}_{n} = \mathbf{m}_{n}(\tau)$ is the contribution to $\mathbf{m}$ in the $n$-th approximation ($|\mathbf{m}_{n}| \sim h^{n}$). Due to the condition $|\mathbf{m}|=1$, there are strong connections between $\mathbf{m}_{n}$ with different $n$. Indeed, using the fact that $\mathbf{m}_{0} = \mathbf{e}_{a}$, the condition $|\mathbf{m}|=1$ can be written as
\begin{equation}
    \sum_{n=1}^{\infty}\mathbf{m}_{n}^{2} +
    2\sum_{n=1}^{\infty}\mathbf{m}_{n}\cdot
    \mathbf{e}_{a} + 2\sum_{n=1}^{\infty}
    \sum_{k=n+1}^{\infty}\mathbf{m}_{n}\cdot
    \mathbf{m}_{k} = 0.
    \label{cond_m=1}
\end{equation}
Since the set of vectors $\mathbf{m}_{n}$, which are introduced instead of the single vector $\mathbf{m}$, is infinite, one may require that the condition (\ref{cond_m=1}) holds in all orders of the perturbation theory, implying that each sum of terms that have the same order equals zero. In this case, for the terms of odd ($n=2p-1$, $p$ is a natural number) and even ($n=2p$) orders one respectively obtains
\begin{equation}
    \mathbf{m}_{2p-1}\cdot \mathbf{e}_{a} =
    - \sum_{l=1}^{p-1}\mathbf{m}_{l} \cdot
    \mathbf{m}_{2p-1-l}
    \label{order_2p-1}
\end{equation}
and
\begin{equation}
    \mathbf{m}_{2p}\cdot \mathbf{e}_{a} =
    - \frac{1}{2}\mathbf{m}_{p}^{2} -
    \sum_{l=1}^{p-1}\mathbf{m}_{l}\cdot
    \mathbf{m}_{2p-l}.
    \label{order_2p}
\end{equation}
Thus, although $\mathbf{m}_{n}$ is determined in the $n$-th step of approximation, the scalar product $\mathbf{m}_{n}\cdot \mathbf{e}_{a}$ can be calculated using $\mathbf{m}_{l}$ obtained in the previous steps (i.e., at $l<n$). This property of $\mathbf{m}_{n}$ plays an important role in our analysis. Note also that for $p=1$ the sums in the right-hand sides of expressions (\ref{order_2p-1}) and (\ref{order_2p}) equal zero, and so
\begin{equation}
    \mathbf{m}_{1}\cdot \mathbf{e}_{a} = 0, \quad
    \mathbf{m}_{2}\cdot \mathbf{e}_{a} =
    - \frac{1}{2}\mathbf{m}_{1}^{2}.
    \label{1-st,2-nd}
\end{equation}

The series representation for the dimensionless effective magnetic field reads
\begin{equation}
    \mathbf{h}_{\mathrm{eff}} = \sum_{n=0}^{\infty}
    \mathbf{h}_{n},
    \label{ser_h}
\end{equation}
where, according to (\ref{h_eff}), (\ref{ser_m}) and (\ref{1-st,2-nd}), $\mathbf{h}_{0} = \mathbf{e}_{a}$, $\mathbf{h}_{1} = \mathbf{H} /H_{a} = h\cos{(\Omega \tau)} \mathbf{e}_{x} + \rho h \sin{(\Omega \tau)} \mathbf{e}_{y}$, and $\mathbf{h}_{n} = {(\mathbf{m}_{n}\cdot \mathbf{e}_{a})} \mathbf{e}_{a}$ at $n\geq 2$. Substituting the series (\ref{ser_m}) and (\ref{ser_h}) into the reduced LLG equation (\ref{red_LLG}) and keeping the terms of the $n$-th order in $h$, we find the following first-order differential equation for $\mathbf{m}_{n}$:
\begin{eqnarray}
    \ell\dot{\mathbf{m}}_{n} &=& -
    \sum_{l=0}^{n}\mathbf{m}_{l} \times
    \mathbf{h}_{n-l} + \alpha\mathbf{h}_{n}
    \nonumber \\
    && - \alpha\sum_{l=0}^{n}\sum_{k=0}^{n-l}
    (\mathbf{m}_{k} \cdot \mathbf{h}_{n-l-k})
    \mathbf{m}_{l}. \quad
    \label{eq_m_n1}
\end{eqnarray}
Finally, by separating the terms with $\mathbf{m}_{n}$, this equation reduces to
\begin{equation}
    \ell\dot{\mathbf{m}}_{n} +
    \alpha \mathbf{m}_{n} + \mathbf{m}_{n}
    \times \mathbf{e}_{a} = \mathbf{f}_{n}
    \label{eq_m_n2}
\end{equation}
($n\geq 1$). Its right-hand side, $\mathbf{f}_{n} = \mathbf{f}_{n}(\tau)$, is given by
\begin{equation}
    \mathbf{f}_{n} = - \!\sum_{l=0}^{n-1}
    \mathbf{m}_{l} \times \mathbf{h}_{n-l}
    + \alpha\mathbf{h}_{n} - \alpha\sum_{
    l=0}^{n-1} \sum_{k=0}^{n-l}(\mathbf{m}
    _{k} \cdot \mathbf{h}_{n-l-k}) \mathbf
    {m}_{l}
    \label{def_f_n}
\end{equation}
and, due to the conditions (\ref{order_2p-1}) and (\ref{order_2p}), it does not depend on $\mathbf{m}_{n}$. Thus, Eq.~(\ref{eq_m_n2}) is linear in $\mathbf{m}_{n}$, and $\mathbf{f}_{n}$ is a given function of $\tau$, which is determined by solving this equation with respect to $\mathbf{m}_{l}$ for $l<n$. In particular, in the first ($n=1$) and second ($n=2$) approximations the definition (\ref{def_f_n}), together with the conditions (\ref{1-st,2-nd}), yields
\begin{equation}
    \mathbf{f}_{1} = \mathbf{h}_{1} \times
    \mathbf{e}_{a} + \alpha\mathbf{h}_{1} -
    \alpha (\mathbf{h}_{1} \cdot \mathbf{e}_{a})
    \mathbf{e}_{a}
    \label{def_h_1}
\end{equation}
and
\begin{eqnarray}
    \mathbf{f}_{2} &=& \mathbf{h}_{1} \times
    \mathbf{m}_{1} - \alpha (\mathbf{h}_{1}
    \cdot \mathbf{m}_{1})\mathbf{e}_{a} -
    \alpha(\mathbf{h}_{1}\cdot \mathbf{e}_{a})
    \mathbf{m}_{1}
    \nonumber \\[4pt]
    && + \,(\alpha/2)\mathbf{m}
    _{1}^{2} \mathbf{e}_{a}.
    \label{def_h_2}
\end{eqnarray}

To write the vector formula (\ref{def_f_n}) in the component form, it is convenient to introduce a right-handed Cartesian coordinate system $x'y'z'$ characterized by the unit vectors $\mathbf{e}_{1}$, $\mathbf{e}_{2}$ and $\mathbf{e}_{a}$. The vector $\mathbf{e}_{a}$ is defined by (\ref{e_a}), and the others may be defined in the following way (see, e.g., Ref.~[\onlinecite{BMS}], p.~162):
\begin{equation}
    \mathbf{e}_{1} = \frac{1}{\sin{\theta_{a}}}
    (\mathbf{e}_{z}\times \mathbf{e}_{a})\times
    \mathbf{e}_{a}, \quad
    \mathbf{e}_{2} = \frac{1}{\sin{\theta_{a}}}
    \mathbf{e}_{z}\times \mathbf{e}_{a}.
    \label{e1,e2}
\end{equation}
Representing in this coordinate system $\mathbf{m}_{n}$ and $\mathbf{f}_{n}$ as
\begin{equation}
    \begin{array}{rcl}
    \mathbf{m}_{n}\!\! &=& \!\!m_{n1}\mathbf{e}_{1}
    + m_{n2}\mathbf{e}_{2} + m_{n3}\mathbf{e}_{a},
    \\ [6pt]
    \mathbf{f}_{n}\!\! &=& \!\!f_{n1}\mathbf{e}_{1}
    + f_{n2}\mathbf{e}_{2} + f_{n3}\mathbf{e}_{a}
    \end{array}
    \label{m,f}
\end{equation}
and taking into account that $m_{n3}$, depending on parity of $n$, is given by (\ref{order_2p-1}) or (\ref{order_2p}), from Eq.~(\ref{eq_m_n2}) for $m_{n1}$ and $m_{n2}$ we obtain a system of equations
\begin{equation}
    \begin{array}{rcl}
    \ell\dot{m}_{n1} + \alpha m_{n1}
    + m_{n2 }\!\! &=& \!\!f_{n1},
    \\ [4pt]
    \ell\dot{m}_{n2} + \alpha m_{n2}
    - m_{n1 }\!\! &=& \!\!f_{n2}.
    \end{array}
    \label{m_n1,m_n2}
\end{equation}

In the steady state, the right-hand sides of these equations are periodic functions of $\tau$, which in the case of even $n$ ($n = 2p-1$) can be written in the matrix form
\begin{equation}
    \begin{pmatrix}
    f_{n1} \\[2pt] f_{n2}
    \end{pmatrix} = \sum_{i=1}^{p}
    \begin{pmatrix}
    q_{ni} & g_{ni} \\[2pt] u_{ni} & v_{ni}
    \end{pmatrix}
    \begin{pmatrix}
    \sin[(2i-1)\Omega\tau] \\[2pt] \cos[(2i-1)
    \Omega\tau]
    \end{pmatrix}\!.
    \label{repr_f1}
\end{equation}
While the matrix elements in (\ref{repr_f1}) have to be determined in the previous steps, the matrix elements in the representation
\begin{equation}
    \begin{pmatrix}
    m_{n1} \\[2pt] m_{n2}
    \end{pmatrix} = \sum_{i=1}^{p}
    \begin{pmatrix}
    a_{ni} & b_{ni} \\[2pt] c_{ni} & d_{ni}
    \end{pmatrix}
    \begin{pmatrix}
    \sin[(2i-1)\Omega\tau] \\[2pt] \cos[(2i-1)
    \Omega\tau]
    \end{pmatrix}
    \label{repr_m1}
\end{equation}
of the steady-state solution of Eqs.~(\ref{m_n1,m_n2}) satisfy the matrix equation
\begin{equation}
    \begin{pmatrix}
    \ell\Omega_{ni} & \alpha & 0 & 1
    \\
    \alpha & -\ell\Omega_{ni} & 1 & 0
    \\
    0 & -1 & \ell\Omega_{ni} & \alpha
    \\
    -1 & 0 & \alpha & -\ell\Omega_{ni}
    \end{pmatrix} \!
    \begin{pmatrix}
    a_{ni} \\ b_{ni} \\ c_{ni} \\ d_{ni}
    \end{pmatrix}
    = \begin{pmatrix}
    g_{ni} \\ q_{ni} \\ v_{ni} \\ u_{ni}
    \end{pmatrix}
    \label{eq_abcd}
\end{equation}
with $\Omega_{ni} = (2i-1)\Omega$. In the case of odd $n$ ($n=2p$), the formulas (\ref{repr_f1}) and (\ref{repr_m1}) should be replaced by
\begin{equation}
    \begin{pmatrix}
    f_{n1} \\[2pt] f_{n2}
    \end{pmatrix} =
    \begin{pmatrix}
    g_{n0} \\[2pt] v_{n0}
    \end{pmatrix} +
    \sum_{i=1}^{p}
    \begin{pmatrix}
    q_{ni} & g_{ni} \\[2pt] u_{ni} & v_{ni}
    \end{pmatrix}
    \begin{pmatrix}
    \sin(2i \Omega\tau) \\[2pt]
    \cos(2i \Omega\tau)
    \end{pmatrix}
    \label{repr_f2}
\end{equation}
and
\begin{equation}
    \begin{pmatrix}
    m_{n1} \\[2pt] m_{n2}
    \end{pmatrix} =
    \begin{pmatrix}
    b_{n0} \\[2pt] d_{n0}
    \end{pmatrix} +
    \sum_{i=1}^{p}
    \begin{pmatrix}
    a_{ni} & b_{ni} \\[2pt] c_{ni} & d_{ni}
    \end{pmatrix}
    \begin{pmatrix}
    \sin(2i\Omega\tau) \\[2pt]
    \cos(2i\Omega\tau)
    \end{pmatrix}\!,
    \label{repr_m2}
\end{equation}
respectively. According to Eqs.~(\ref{m_n1,m_n2}), the parameters $b_{n0}$ and $d_{n0}$, i.e., the time-independent parts of $m_{n1}$ and $m_{n2}$, are determined by the equations
\begin{equation}
    \begin{array}{rcl}
    \alpha b_{n0} + d_{n0}\!\! &=& \!\!g_{n0},
    \\ [4pt]
    - b_{n0} + \alpha d_{n0}\!\! &=& \!\!v_{n0}
    \end{array}
    \label{eq_b0,d0}
\end{equation}
(the parameters $g_{n0}$ and $v_{n0}$ are assumed to be known), and the matrix elements in (\ref{repr_m2}) satisfy the same equation (\ref{eq_abcd}) with $\Omega_{ni} = 2i\Omega$.

The solution of Eqs.~(\ref{eq_b0,d0}) is given by
\begin{equation}
    b_{n0} = \frac{1}{\ell}(\alpha g_{n0} - v_{n0}),
    \quad
    d_{n0} = \frac{1}{\ell}(g_{n0} + \alpha v_{n0}),
    \label{b0,d0}
\end{equation}
and the solution of the matrix equation (\ref{eq_abcd}) reads
\begin{eqnarray}
    a_{ni} \!&=&\! \displaystyle\frac{1}
    {\ell\Delta_{ni}} \big[- \Omega_{ni}(1 - \alpha^{2}-
    \ell^{2}\Omega_{ni}^{2})g_{ni}
    + \alpha (1 + \ell\Omega_{ni}^{2})q_{ni}
    \nonumber\\
    && \! - \,2\alpha \Omega_{ni}v_{ni} -
    (1 - \ell\Omega_{ni}^{2})u_{ni} \big],
    \nonumber\\[2pt]
    b_{ni} \!&=&\! \displaystyle \frac{1}
    {\ell\Delta_{ni}}\big[\alpha (1 + \ell\Omega_{ni}^{2})
    g_{ni} + \Omega_{ni}(1 - \alpha^{2} - \ell^{2}
    \Omega_{ni}^{2})q_{ni}
    \nonumber\\
    && \!- \,(1 - \ell\Omega_{ni}^{2})v_{ni} +
    2\alpha \Omega_{ni}u_{ni}\big],
    \nonumber\\[2pt]
    c_{ni} \!&=&\! \displaystyle \frac{1}
    {\ell\Delta_{ni}} \big[2\alpha \Omega_{ni}g_{ni} +
    (1 - \ell\Omega_{ni}^{2})q_{ni}
    \nonumber\\
    && \!- \,\Omega_{ni}(1 - \alpha^{2}- \ell^{2}
    \Omega_{ni}^{2})v_{ni} + \alpha (1 + \ell
    \Omega_{ni}^{2})u_{ni}\big],
    \nonumber\\[2pt]
    d_{ni} \!&=&\! \displaystyle \frac{1}
    {\ell\Delta_{ni}} \big[(1 - \ell\Omega_{ni}^{2})
    g_{ni} - 2\alpha \Omega_{ni}q_{ni}
    \nonumber\\
    && \!+ \,\alpha (1 + \ell\Omega_{ni}^{2})
    v_{ni} + \Omega_{ni}(1 - \alpha^{2}- \ell^{2}
    \Omega_{ni}^{2})u_{ni}\big].
    \nonumber\\
    \label{abcd_ni}
\end{eqnarray}
Here, $\Omega_{ni} = (2i-1) \Omega$ if $n = 2p-1$, $\Omega_{ni} = 2i \Omega$ if $n = 2p$, $i = \overline{1,p}$, and
\begin{equation}
    \Delta_{ni} = (1 - \ell \Omega_{ni}^{2})^{2}
    + 4\alpha^{2}\Omega_{ni}^{2}.
    \label{Delta}
\end{equation}

Below, we consider in more detail the first- and second-order approximations and discuss qualitatively the role of higher-order terms in the perturbation expansion of the steady-state magnetization.

\subsection{First-order approximation}
\label{1-st}

In this approximation, the contribution to the steady-state solution of the LLG equation, $\mathbf{m}_{1}$, can easily be found from the general expressions (\ref{abcd_ni}), in which  $n=i=p=1$ and $\Omega_{ni} = \Omega$. Indeed, rewriting the representation (\ref{repr_f1}) in the form
\begin{equation}
    \begin{array}{rcl}
    f_{11}\!\! &=& \!\! q_{11}\sin(\Omega\tau) +
    g_{11} \cos(\Omega\tau),
    \\ [4pt]
    f_{12}\!\! &=& \!\! u_{11}\sin(\Omega\tau) +
    v_{11} \cos(\Omega\tau)
    \end{array}
    \label{f11}
\end{equation}
and using the vector formula (\ref{def_h_1}) together with the definitions (\ref{e_a}) and (\ref{e1,e2}), one straightforwardly gets
\begin{equation}
    \begin{alignedat}{2}
    q_{11} &= \rho h(\kappa_{a} + \alpha\lambda_{a}
    \delta_{a}), & \quad
    g_{11} &= -h(\delta_{a} - \alpha\lambda_{a}
    \kappa_{a}),
    \\[2pt]
    u_{11} &= -\rho h(\lambda_{a}\delta_{a} -
    \alpha\kappa_{a}), & \quad
    v_{11} &= -h(\lambda_{a} \kappa_{a} +
    \alpha\delta_{a}),
    \label{qguv_11}
    \end{alignedat}
\end{equation}
where, for the sake of brevity here and in the following, we have introduced  the notations
\begin{equation}
    \begin{alignedat}{2}
    \kappa_{a} &= \cos\varphi_{a}, & \quad
    \delta_{a} &= \sin\varphi_{a},
    \\[2pt]
    \lambda_{a} &= \cos\theta_{a}, & \quad
    \chi_{a} &= \sin\theta_{a}.
    \label{defs1}
    \end{alignedat}
\end{equation}
Substituting the representation coefficients (\ref{qguv_11}) into expressions (\ref{abcd_ni}), we eventually find
\begin{eqnarray}
    a_{11} \!&=&\! \displaystyle\frac{h}
    {\Delta_{11}}\big[2\alpha \rho\Omega^{2}
    \kappa_{a} + \alpha\Omega(1 + \ell
    \Omega^{2}) \lambda_{a}\kappa_{a}
    \nonumber\\
    && \!+ \,\Omega(1-\ell\Omega^{2})
    \delta_{a} + \rho (1- 2\Omega^{2} +
    \ell \Omega^{2})\lambda_{a}\delta_{a}\big],
    \nonumber\\[2pt]
    b_{11} \!&=&\! \displaystyle \frac{h}
    {\Delta_{11}} \big[\rho\Omega(1-\ell
    \Omega^{2}) \kappa_{a} + (1 - 2\Omega^{2}
    +\ell \Omega^{2})\lambda_{a} \kappa_{a}
    \nonumber\\
    && \!- \,2\alpha \Omega^{2}\delta_{a} -
    \alpha \rho \Omega(1+\ell\Omega^{2})
    \lambda_{a}\delta_{a}\big],\quad
    \nonumber\\[2pt]
    c_{11} \!&=&\! \displaystyle\frac{h}
    {\Delta_{11}} \big[\rho(1-2\Omega^{2} +
    \ell \Omega^{2}) \kappa_{a} + \Omega(1 -
    \ell\Omega^{2})\lambda_{a}\kappa_{a}
    \nonumber\\
    && \!- \,\alpha\Omega(1+\ell\Omega^{2})
    \delta_{a}  - 2\alpha\rho\Omega^{2}\lambda_{a}
    \delta_{a}\big],
    \nonumber\\[2pt]
    d_{11} \!&=&\! \displaystyle\frac{h}
    {\Delta_{11}}\big[-\alpha \rho\Omega(1+
    \ell\Omega^{2})\kappa_{a} - 2\alpha\Omega^{2}
    \lambda_{a}\kappa_{a}
    \nonumber\\
    && \!- \, (1-2\Omega^{2} + \ell\Omega^{2})
    \delta_{a} - \rho\Omega (1-\ell\Omega^{2})
    \lambda_{a}\delta_{a}\big].
    \nonumber\\
    \label{abcd_11}
\end{eqnarray}

Since, according to (\ref{Delta}),
\begin{equation}
    \Delta_{11} = (1 - \ell \Omega^{2})^{2}
    + 4\alpha^{2}\Omega^{2}
    \label{Delta11}
\end{equation}
and the condition $\alpha\ll 1$ usually holds, the frequency dependence of the first-order contribution to the steady-state reduced magnetization,
\begin{eqnarray}
    \mathbf{m}_{1} &=& (a_{11}
    \mathbf{e}_{1} + c_{11}\mathbf{e}_{2})
    \sin(\Omega \tau)
    \nonumber \\[4pt]
    && + \,(b_{11}  \mathbf{e}_{1} + d_{11}
    \mathbf{e}_{2}) \cos(\Omega \tau),
    \label{m1}
\end{eqnarray}
exhibits, in general, a resonant behavior near the ferromagnetic resonance frequency $(\Omega=1)$. Note, however, that in some particular cases, e.g., if $\theta_{a} = \varphi_{a} = 0$ and $\rho = -1$ or if $\theta_{a} = \pi/2$, $\varphi_{0} = 0$ and $\rho = 0$, the resonance does not exist. In the first case, the physical reason is that the direction of the magnetic field rotation is opposite to the direction of the natural precession of the nanoparticle magnetization.\cite{DLH} In contrast, in the second case, the reason is that the magnetic field, whose direction is parallel to the anisotropy axis of the nanoparticles, does not induce the magnetization dynamics in this approximation. It is important to emphasize that, since the characteristics of the considered system are averaged over all the directions of the unit vector $\mathbf{e}_{a}$, which is assumed to be uniformly distributed over the sphere, there is no contribution from nanoparticles with fixed $\theta_{a}$ and $\varphi_{a}$.

\subsection{Second-order approximation}
\label{2-nd}

The representation (\ref{repr_f2}) corresponds to  the second-order approximation if $n=2$ and $i=p=1$, yielding
\begin{equation}
    \begin{array}{rcl}
    f_{21}\!\! &=& \!\! g_{20} + q_{21}\sin(2
    \Omega\tau) + g_{21} \cos(2\Omega\tau),
    \\ [4pt]
    f_{22}\!\! &=& \!\! v_{20} + u_{21}\sin(2
    \Omega\tau) + v_{21} \cos(2\Omega\tau).
    \end{array}
    \label{f21}
\end{equation}
From this, using (\ref{def_h_2}) and (\ref{m1}), we find
\begin{equation}
    \begin{array}{rcl}
    g_{20}\!\! &=& \!\! \displaystyle -\frac{h}{2}
    \big[(d_{11} + \alpha b_{11})\kappa_{a} + \rho
    (c_{11} + \alpha a_{11})\delta_{a}\big]\chi_{a},
    \\ [8pt]
    v_{20}\!\! &=& \!\! \displaystyle \frac{h}{2}
    \big[(b_{11} - \alpha d_{11})\kappa_{a} + \rho
    (a_{11} - \alpha c_{11})\delta_{a}\big]\chi_{a}
    \end{array}
    \label{gv0}
\end{equation}
and
\begin{equation}
    \begin{array}{rcl}
    q_{21}\!\! &=& \!\! \displaystyle -\frac{h}{2}
    \big[(c_{11} + \alpha a_{11})\kappa_{a} + \rho
    (d_{11} + \alpha b_{11})\delta_{a}\big]\chi_{a},
    \\ [8pt]
    g_{21}\!\! &=& \!\! \displaystyle -\frac{h}{2}
    \big[(d_{11} + \alpha b_{11})\kappa_{a} - \rho
    (c_{11} + \alpha a_{11})\delta_{a}\big]\chi_{a},
    \\ [8pt]
    u_{21}\!\! &=& \!\! \displaystyle \frac{h}{2}
    \big[(a_{11} - \alpha c_{11})\kappa_{a} + \rho
    (b_{11} - \alpha d_{11})\delta_{a}\big]\chi_{a},
    \\ [8pt]
    v_{21}\!\! &=& \!\! \displaystyle \frac{h}{2}
    \big[(b_{11} - \alpha d_{11})\kappa_{a} - \rho
    (a_{11} - \alpha c_{11})\delta_{a}\big]\chi_{a}.
    \end{array}
    \label{qguv2}
\end{equation}
Then, rewriting the representation (\ref{repr_m2}) in the form
\begin{equation}
    \begin{array}{rcl}
    m_{21}\!\! &=& \!\! b_{20} + a_{21} \sin(2\Omega\tau) +
    b_{21}\cos(2\Omega\tau),
    \\ [4pt]
    m_{22}\!\! &=& \!\! d_{20} + c_{21} \sin(2\Omega\tau) +
    d_{21}\cos(2\Omega\tau)
    \end{array}
    \label{m21}
\end{equation}
and using (\ref{gv0}), from (\ref{b0,d0}) one immediately obtains
\begin{equation}
    \begin{array}{rcl}
    b_{20}\!\! &=& \!\! \displaystyle -\frac{h}{2}
    (b_{11}\kappa_{a} + \rho a_{11}\delta_{a})\chi_{a},
    \\ [6pt]
    d_{20}\!\! &=& \!\! \displaystyle -\frac{h}{2}
    (d_{11}\kappa_{a} + \rho c_{11}\delta_{a})\chi_{a}.
    \end{array}
    \label{bd_20}
\end{equation}

Finally, substituting the coefficients (\ref{qguv2}) into the general expressions (\ref{abcd_ni}), we get
\begin{eqnarray}
    a_{21} \!&=&\! \displaystyle\frac{h\chi_{a}}
    {2\Delta_{21}}\big[ -(1- 8\Omega^{2} + 4\ell
    \Omega^{2})A -2\alpha \Omega(1+ 4\ell\Omega^{2})B
    \nonumber\\
    && \! -\,8\alpha \Omega^{2}C + 2\Omega(1 -4\ell
    \Omega^{2})D\big],
    \nonumber\\[2pt]
    b_{21} \!&=&\! \displaystyle \frac{h\chi_{a}}
    {2\Delta_{21}} \big[2\alpha\Omega(1+ 4\ell
    \Omega^{2})A - (1- 8\Omega^{2}+ 4\ell\Omega^{2})B
    \nonumber\\
    && \!- \,2\Omega(1 -4\ell\Omega^{2})C - 8\alpha
    \Omega^{2}D\big],
    \nonumber\\[2pt]
    c_{21} \!&=&\! \displaystyle\frac{h\chi_{a}}
    {2\Delta_{21}} \big[8\alpha \Omega^{2}A - 2\Omega
    (1 -4\ell\Omega^{2})B
    \nonumber\\
    && \!- \,(1 -8\Omega^{2} + 4\ell\Omega^{2})C -
    2\alpha\Omega(1+ 4\ell\Omega^{2})D \big],
    \nonumber\\[2pt]
    d_{21} \!&=&\! \displaystyle\frac{h\chi_{a}}
    {2\Delta_{21}}\big[2\Omega (1 -4\ell\Omega^{2})A +
    8\alpha \Omega^{2}B
    \nonumber\\
    && \!+ \, 2\alpha\Omega(1 +4\ell\Omega^{2})C -
    (1 -8\Omega^{2} + 4\ell\Omega^{2})D\big],
    \nonumber\\
    \label{abcd_21}
\end{eqnarray}
where
\begin{equation}
    \begin{alignedat}{2}
    A &= a_{11}\kappa_{a} + \rho b_{11}\delta_{a},
    & \quad
    B &= b_{11}\kappa_{a} - \rho a_{11}\delta_{a},
    \\[2pt]
    C &= c_{11}\kappa_{a} + \rho d_{11}\delta_{a},
    & \quad
    D &= d_{11}\kappa_{a} - \rho c_{11}\delta_{a},
    \label{ABCD}
    \end{alignedat}
\end{equation}
and, according to the definition (\ref{Delta}),
\begin{equation}
    \Delta_{21} = (1 - 4\ell \Omega^{2})^{2}
    + 16\alpha^{2}\Omega^{2}.
    \label{Delta21}
\end{equation}

Thus, the second-order contribution to the steady-state reduced magnetization is given by
\begin{eqnarray}
    \mathbf{m}_{2} &=& b_{20}\mathbf{e}_{1} +
    d_{20}\mathbf{e}_{2} + (a_{21}\mathbf{e}_{1}
    + c_{21}\mathbf{e}_{2}) \sin(2\Omega \tau)
    \nonumber \\[2pt]
    && + \,(b_{21}  \mathbf{e}_{1} + d_{21}
    \mathbf{e}_{2}) \cos(2\Omega \tau) - \frac{1}{2}
    \mathbf{m}_{1}^{2}\mathbf{e}_{a}. \quad
    \label{m2}
\end{eqnarray}
This contribution, in contrast to the first-order one, has a resonant dependence on the reduced frequency $\Omega$ not only in the vicinity of the first-order resonance $(\Omega=1)$ but, as it follows from (\ref{abcd_21}) and (\ref{Delta21}), also in the vicinity of the second-order resonance $(\Omega=1/2)$. It should be noted that for nanoparticles, whose anisotropy axes are perpendicular to the polarization plane (when $\chi_{a} = 0$), this effect does not exist.

\section{AVERAGE MAGNETIZATION}
\label{Magn}

Now, using the above results of the perturbation theory, we determine the average value of the reduced nanoparticle magnetization, $\langle\overline{ \mathbf{m}} \rangle$, in the quadratic approximation. Since in this case $\mathbf{m} = \mathbf{e}_{a} + \mathbf{m}_{1} + \mathbf{m}_{2}$ with $\mathbf{m}_{1}$ and $\mathbf{m}_{2}$ given by (\ref{m1}) and (\ref{m2}), respectively, the time averaging of $\mathbf{m}$ yields
\begin{equation}
    \overline{\mathbf{m}} = b_{20}\mathbf{e}_{1}
    + d_{20} \mathbf{e}_{2} + \big(1 - \tfrac{1}{2}
    \overline{\mathbf{m}_{1}^{2}}\big)\mathbf{e}_{a}.
    \label{av_m}
\end{equation}
From this and from the definitions (\ref{e1,e2}) and (\ref{e_a}) of the unit vectors $\mathbf{e}_{1}$, $\mathbf{e}_{2}$ and $\mathbf{e}_{a}$, the Cartesian components of $\overline{\mathbf{m}}$ can be written in the form
\begin{subequations}
    \begin{align}
    \overline{m_{x}} &= b_{20}
    \lambda_{a}\kappa_{a} - d_{20}\delta_{a}
    + \big(1 - \tfrac{1}{2}\overline{\mathbf{m}_{1}^{2}}
    \big) \chi_{a}\kappa_{a},\label{m_x} \\[3pt]
    \overline{m_{y}} &= b_{20}
    \lambda_{a}\delta_{a} + d_{20}\kappa_{a}
    + \big(1 - \tfrac{1}{2}\overline{\mathbf{m}_{1}^{2}}
    \big) \chi_{a}\delta_{a},\label{m_y} \\[3pt]
    \overline{m_{z}} &= - b_{20}
    \chi_{a} + \big(1 - \tfrac{1}{2}\overline{
    \mathbf{m}_{1}^{2}} \big) \lambda_{a}.\label{m_z}
    \end{align}
    \label{m_xyz}
\end{subequations}
Then, we average these components over the angles $\theta_{a}$ and  $\varphi_{a}$ distributed with the probability density (\ref{P}).  Taking into account that $\langle F(\theta_{a}, \varphi_{a})\rangle = 0$ if  $F(\pi - \theta_{a}, \varphi_{a}) = -F(\theta_{a}, \varphi_{a})$ or $F(\theta_{a}, \pi + \varphi_{a}) = -F(\theta_{a}, \varphi_{a})$, one can make sure that all averages in the right-hand sides of (\ref{m_x}) and (\ref{m_y}) are equal to zero, and so
\begin{equation}
    \langle \overline{m_{x}} \rangle =
    \langle \overline{m_{y}} \rangle =0.
    \label{<mx_my>}
\end{equation}

In contrast,  with the exception of $\langle \lambda_{a} \rangle = 0$, the other averages in the right-hand side of (\ref{m_z}), i.e., $\langle b_{20} \chi_{a} \rangle$ and $\langle \overline{\mathbf{m}_{1}^{2}} \lambda_{a} \rangle$, are not equal to zero. Indeed, using the previously derived results (\ref{bd_20}) and (\ref{abcd_11}) together with the conditions $\langle \chi_{a}^{2} \kappa_{a}^{2} \rangle = \langle \chi_{a}^{2} \delta_{a}^{2} \rangle = 1/3$, which can be verified directly from the definition (\ref{mean2}), we obtain
\begin{equation}
    \langle b_{20} \chi_{a} \rangle = -
    \frac{\rho h^{2}\Omega} {3\Delta_{11}}
    (1 - \ell \Omega^{2}).
    \label{<b20chi>}
\end{equation}
Similarly, taking into account that
\begin{equation}
    \overline{\mathbf{m}_{1}^{2}} = \frac{1}{2}
    \big(a_{11}^{2} + b_{11}^{2} + c_{11}^{2} +
    d_{11}^{2}\big)
    \label{av_m1}
\end{equation}
and $\langle \lambda_{a}^{2} \kappa_{a}^{2} \rangle = \langle \lambda_{a}^{2} \delta_{a}^{2} \rangle = 1/6$, one can show that
\begin{eqnarray}
    \langle \overline{\mathbf{m}_{1}^{2}}
    \lambda_{a}\rangle &=& \frac{2\rho h^{2}
    \Omega}{3\Delta_{11}^{2}}\big[ (1 - \ell
    \Omega^{2})(1 - 2\Omega^{2} + \ell\Omega^{2})
    \nonumber \\[2pt]
    && + \,2\alpha^{2} \Omega^{2} (1 + \ell
    \Omega^{2})\big].
    \label{<m_lambda>}
\end{eqnarray}
Finally, since $\langle \overline{m_{z}} \rangle = - \langle b_{20} \chi_{a} \rangle - \langle \overline{\mathbf{m}_{1}^{2}} \lambda_{a}\rangle/2$, from (\ref{<b20chi>}) and (\ref{<m_lambda>}) it follows that $\langle \overline{m_{z}} \rangle = -\rho \ell h^{2}\Omega^{3}/ (3\Delta_{11})$ or, with the notation (\ref{Delta11}),
\begin{equation}
    \langle \overline{m_{z}} \rangle = -
    \frac{1}{3}\rho \ell h^{2} \frac{
    \Omega^{3}}{(1 - \ell\Omega^{2})^{2} +
    4\alpha^{2} \Omega^{2}}.
    \label{<mz>}
\end{equation}

Thus, the elliptically polarized magnetic field (\ref{H}), which has no constant components, magnetizes the considered systems of ferromagnetic nanoparticles. Since these systems are characterized by the uniform distribution of easy axis directions, there is no net magnetization without this field. In its presence, the direction of induced magnetization is perpendicular to the polarization plane and depends on the direction of the magnetic field rotation (i.e., on the sign of $\rho$). The phenomenon of induced magnetization has a purely dynamical origin: according to the LLG equation (\ref{red_LLG}), the forced dynamics of the reduced magnetization $\mathbf{m}$ in nanoparticles characterized by the vectors $\mathbf{e}_{a}$ and $-\mathbf{e}_{a}$ is quite different. It is worthwhile to recall that the induced magnetization (\ref{<mz>}) is the second-order effect. Its main feature is that $\langle \overline{m_{z}} \rangle$ in the vicinity of the point $\Omega =1$ depends on $\Omega$ in a resonant manner ($\max |\langle \overline{m_{z}} \rangle| \simeq |\rho| h^{2}/ (12\alpha^{2})$ as $\alpha \ll 1$). It should also be noted that the linearly polarized magnetic field (when $\rho =0$) does not magnetize the considered systems of ferromagnetic nanoparticles.

\begin{figure}
    \centering
    \includegraphics[totalheight=5cm]{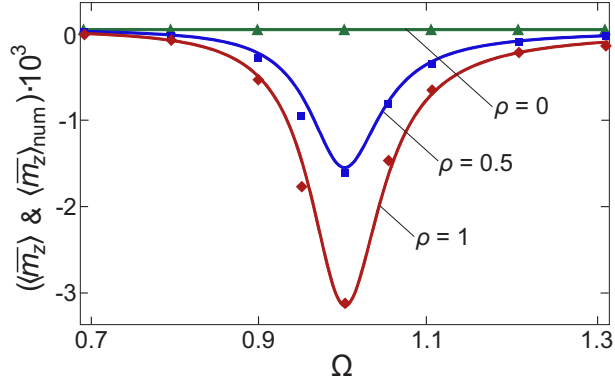}
    \caption{\label{fig1} (Color online) Frequency
    dependence of the $z$ component of the reduced
    magnetization. The solid lines represent the
    theoretical result (\ref{<mz>}), and the
    numerical results for $\langle \overline {m_{z}}
    \rangle_{\mathrm{num}}$ are shown by the
    symbols (their size exceeds the error bars). The
    theoretical and numerical results are presented
    for the circularly ($\rho = 1$), elliptically
    ($\rho = 0.5$), and linearly ($\rho =0$) polarized
    magnetic field of amplitude $h=0.01$.}
\end{figure}

In order to verify the theoretical results, we numerically determined the steady-state solution $\mathbf{m}^{(l)}$ ($l= \overline{1,N}$) of Eq.~(\ref{red_LLG}) (we take $\alpha = 0.05$ in all our numerical calculations) for $N= 2 \cdot 10^{3}$ nanoparticles, whose easy axis directions are distributed according to the probability density (\ref{P}). Then, calculating the average reduced magnetization as $\langle \overline{\mathbf{m}} \rangle_{\mathrm{num}} = (1/N \mathcal {T}) \sum_{l=1}^{N} \int_{0}^{\mathcal {T}} \mathbf{m}^{(l)} d\tau$, we made sure that the Cartesian components of $\langle \overline{\mathbf{m}} \rangle_{\mathrm{num}}$ are in very good agreement with those predicted in (\ref{<mx_my>}) and (\ref{<mz>}), if the reduced magnetic field amplitude $h$ is small enough. For illustration, in Fig.~\ref{fig1} we show the dependence of $\langle \overline{m_{z}} \rangle$ and $\langle \overline{m_{z}} \rangle_{\mathrm{num}}$ on the reduced frequency for different polarizations of the external magnetic field of relatively small amplitude.

\begin{figure}
    \centering
    \includegraphics[totalheight=4.5cm]{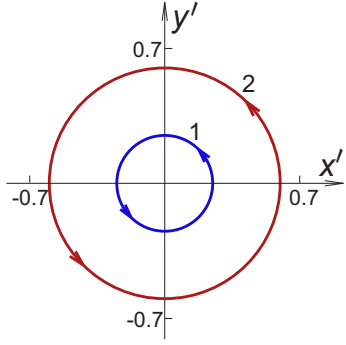}
    \caption{\label{fig2} (Color online) Projection
    of the steady-state trajectories of the reduced
    magnetization on the $x'y'$ plane just below (1)
    and just above (2) the transition frequency.
    The trajectories are obtained from the numerical
    solution of Eq.~(\ref{red_LLG}) for the nanoparticle,
    whose easy axis direction is characterized by the
    angles $\theta_{a} = \pi/3$ and $\varphi_{a} = 0$,
    and the reduced magnetic field of amplitude $h =
    0.05$ has the circular polarization ($\rho =1$).
    In this case, the nanoparticle magnetization rotates
    in the counterclockwise direction (indicated by the
    arrows), the trajectories are almost circular, and
    the transition frequency is approximately given by
    $\Omega_{\mathrm{tr}}\simeq 0.84$.}
\end{figure}

\begin{figure}
    \centering
    \includegraphics[totalheight=5cm]{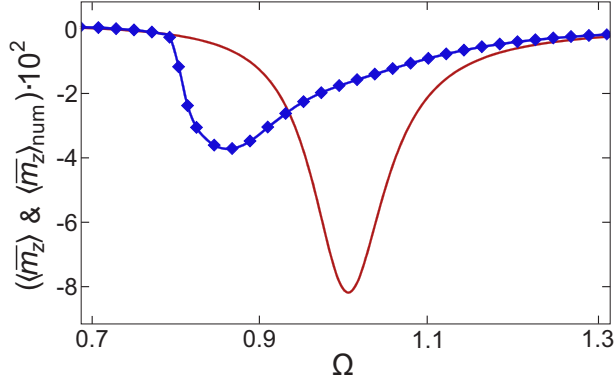}
    \caption{\label{fig3} (Color online) Frequency
    dependence of the $z$ component of the reduced
    magnetization for $h= 0.05$ and $\rho =1$. The
    theoretical ($\langle \overline {m_{z}} \rangle$)
    and numerical ($\langle \overline {m_{z}}\rangle
    _{\mathrm{num}}$) results are shown by the solid
    and symbol lines, respectively. The difference
    between them is caused by the transitions in the
    magnetization dynamics similar to that shown in
    Fig.~\ref{fig2} (these transitions in nanoparticles
    with different easy axis directions occur at
    different frequencies).}
\end{figure}

The difference between $\langle \overline {m_{z}} \rangle$ and $\langle \overline {m_{z}} \rangle_{\mathrm{num}}$ as functions of $\Omega$ appears and grows with increasing $h$. There are a few reasons for this. One of them is that the  magnetization of some nanoparticles, depending on the direction of their easy axes, can transit to a new steady state with increasing $\Omega$, if the reduced amplitude $h$ is not too small. In such a case, the transition occurs at $\Omega = \Omega_{\mathrm{tr}}$, where $\Omega_{ \mathrm{tr}} (<1)$ is the transition frequency, and is accompanied by an abrupt change in the steady-state trajectory of the reduced magnetization, see Fig.~\ref{fig2}. While the steady-state period $\mathcal {T}$ is the same just below and just above the transition frequency, the switching of the steady state leads to a strong change of the $z$ component of the reduced magnetization at $\Omega = \Omega_{ \mathrm{tr}}$ (because $|\mathbf{m}^{(l)}| =1$). As shown in Fig.~\ref{fig3}, due to the existence of the transition frequency $\Omega_{ \mathrm{tr}}$ and its slow dependence on easy axis directions, the frequency dependence of $\langle \overline {m_{z}} \rangle_{ \mathrm {num}}$ qualitatively differs from the theoretical result (\ref{<mz>}). With a slight increase of $h$, the peak of $|\langle \overline {m_{z}} \rangle_{\mathrm {num}}|$ is shifted to lower frequencies and its maximum value decreases, in contrast to the $\Omega$- and $h$-dependence of $|\langle \overline {m_{z}} \rangle|$. At the same time, the condition $\langle \overline {m_{z}} \rangle|_{-\rho} = -\langle \overline {m_{z}} \rangle|_{\rho}$, which follows from (\ref{<mz>}), holds for $\langle \overline {m_{z}} \rangle_{\mathrm {num}}$ as well. We note also that a little difference between $\langle \overline {m_{z}} \rangle$ and $\langle \overline {m_{z}} \rangle_{\mathrm {num}}$ at $h= 0.01$ (see Fig.~\ref{fig1}, $\Omega \simeq 0.95$) arises from the fact that in this case there is a small fraction of nanoparticles in which the transition to a new steady state still occurs. This fraction decreases with decreasing $h$ and, e.g., at $h= 0.005$ the above difference practically vanishes.

The second reason is that the role of the higher-order terms $\mathbf{m}_{n}$ ($n \geq 3$) in the expansion of the reduced magnetization $\mathbf{m}$, which are neglected in $\langle \overline {m_{z}} \rangle$, grows with increasing $h$. Since, according to (\ref{Delta}), these terms depend on $\Omega$ in a resonant way not only near the point $\Omega = 1$ (for example, the resonant behavior of $\mathbf{m}_{n}$ with $n =2p$ occurs in the vicinity of the points $\Omega = 1/(2i)$, $i= \overline{1,p}$), the frequency dependence of $\mathbf{m}$ in the $n$-th order approximation can strongly differ from that obtained in the second-order approximation. Although the average of $\mathbf{m}$ may eliminate some of resonances (in particular, $\langle \overline {m_{z}} \rangle$, in contrast to $m_{z}$, has no resonance at $\Omega = 1/2$ in the second-order approximation), we can expect that, in general, $\langle \overline {m_{z}} \rangle$ as a function of $\Omega$ behaves in a qualitatively different way in the higher-order and second-order approximations.

\begin{figure}
    \centering
    \includegraphics[totalheight=5cm]{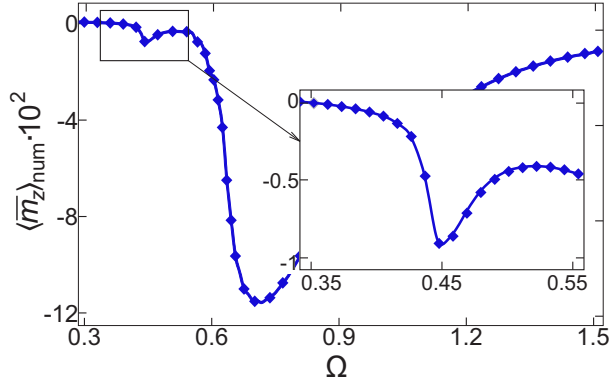}
    \caption{\label{fig4} (Color online) Dependence
    of $\langle \overline {m_{z}} \rangle_{ \mathrm
    {num}}$ on the reduced frequency $\Omega$ for
    $h= 0.14$ and $\rho =1$. Insert: the same
    dependence in the vicinity of the second-order
    resonance.}
\end{figure}

To demonstrate this explicitly, let us first consider the third-order approximation in $h$, when $\mathbf{m} = \mathbf{e}_{a} + \sum_{n=1}^{3} \mathbf{m}_{n}$. Using (\ref{order_2p-1}) and (\ref{repr_m1}), we find $\overline {\mathbf{m}_{3}} \cdot \mathbf{e}_{a} = - \overline {\mathbf{m}_{1} \cdot \mathbf{m}_{2}} = 0$ and $\overline {\mathbf{m}_{3}} \cdot \mathbf{e}_{1} = \overline {\mathbf{m}_{3}} \cdot \mathbf{e}_{2} = 0$, i.e., $\overline {\mathbf{m}_{3}} = 0$. This means that $\langle \overline {\mathbf{m}} \rangle  = \langle \overline {\mathbf{m}_{2}} \rangle$, and thus the formula (\ref{<mz>}) holds also in the third-order approximation. In contrast, in the fourth-order approximation in $h$, when $\mathbf{m} = \mathbf{e}_{a} + \sum_{n=1}^{4} \mathbf{m}_{n}$, we have $\langle \overline {\mathbf{m}} \rangle  = \langle \overline {\mathbf{m}_{2}} \rangle + \langle \overline {\mathbf{m}_{4}} \rangle$, where, according to (\ref{order_2p}) and (\ref{repr_m2}),
\begin{equation}
    \langle \overline {\mathbf{m}_{4}} \rangle =
    - \frac{1}{2} \langle \overline {\mathbf{m}
    _{2}^{2}}\, \mathbf{e}_{a} \rangle - \langle
    (\overline {\mathbf{m}_{1} \cdot \mathbf{m}
    _{3}}) \mathbf{e}_{a} \rangle.
    \label{m_4}
\end{equation}
The resonance at $\Omega =1/3$, which is associated with the frequency dependence of $\mathbf{m}_{3}$, is eliminated by the time averaging, and the last term in (\ref{m_4}), if it is nonzero, resonantly depends on $\Omega$ only in the vicinity of the point $\Omega =1$. Therefore, since this term is of the order of $h^{4}$, it does not change qualitatively the second-order result (\ref{<mz>}). As to the first term in the right-hand side of (\ref{m_4}), it, according to (\ref{bd_20})--(\ref{m2}), exhibits a resonant behavior in the vicinity of two reduced frequencies $\Omega =1/2$ and $\Omega =1$. This term is also of the order of $h^{4}$, but, in general, it can be neglected everywhere except in the vicinity of the point $\Omega =1/2$. In this frequency domain, one may expect that the $z$ component of the first term, $-\langle \overline {\mathbf{m}_{2}^{2}} \lambda_{a} \rangle/2$, exceeds $\langle \overline {m_{z}} \rangle$, if the reduced magnetic field amplitude $h$ is not too small. It is clear from the previous results that, depending on $h$, the $z$ component of $\langle \overline {\mathbf{m}} \rangle$ in the $2p$-th order approximation can exhibit a resonant behavior in the vicinity of the subharmonic frequencies $\Omega = 1/i$ with $i= \overline{1,p}$ (the resonant frequency $\Omega = 1/i$ corresponds to the $i$-th order resonance). The frequency dependence of $\langle \overline {m_{z}} \rangle_{ \mathrm {num}}$, illustrating the role of the second-order resonance, is shown in Fig.~\ref{fig4}. For the same reason as in Fig.~\ref{fig3}, the local minimum of $\langle \overline {m_{z}} \rangle_{ \mathrm {num}}$ is shifted (with respect to the analytical result $\Omega = 1/2$) to lower frequencies.

\begin{figure}
    \centering
    \includegraphics[totalheight=5cm]{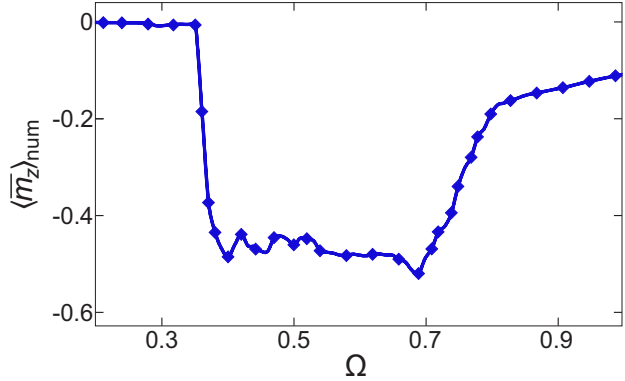}
    \caption{\label{fig5} (Color online) Frequency
    dependence of $\langle \overline {m_{z}} \rangle_{
    \mathrm{num}}$ for $h= 0.25$ and $\rho =1$. The
    nonmonotonic behavior of $\langle \overline
    {m_{z}} \rangle_{\mathrm{num}}$, which occurs
    in the interval $(0.4, 0.7)$, results from
    magnetization switching in some nanoparticles;
    the width of this interval grows with increasing
    $h$.}
\end{figure}

\begin{figure}
    \centering
    \includegraphics[totalheight=5cm]{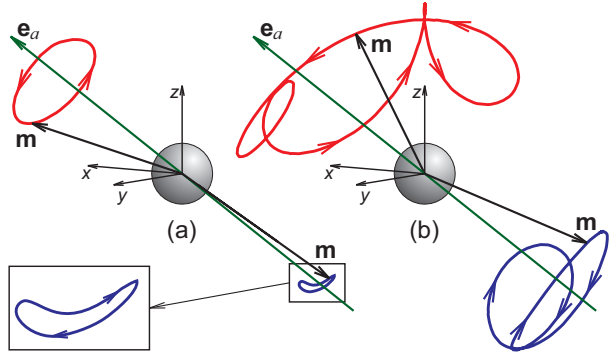}
    \caption{\label{fig6} (Color online) Steady-state
    trajectories of the reduced magnetization
    $\mathbf{m}$ for the nanoparticle with $\theta_{a}
    = \pi/3$ and $\varphi_{a} = 0$ driven by the
    circularly polarized magnetic field ($\rho =1$) of
    amplitude $h = 0.14$ (a) and $h=0.25$ (b). The
    upper/lower trajectories correspond to the field
    frequencies just before/after the switching transition.
    At $h = 0.14$ the upper trajectory corresponds to
    $\Omega = 0.6$, the lower one to $\Omega = 0.601$,
    $\Omega_{\mathrm{tr}} \in (0.6, 0.601)$, and the
    magnetization precession in both states occurs with
    the field frequency $\Omega$. In contrast, at $h =
    0.25$ the upper trajectory corresponds to $\Omega =
    0.485$, the lower one to $\Omega = 0.486$, $\Omega_{
    \mathrm{tr}} \in (0.485, 0.486)$, and, while the
    magnetization precession in the lower state occurs
    with the field frequency, the frequency of precession
    in the upper state is two times less than $\Omega$.}
\end{figure}

With further increasing $h$, the magnetization dynamics becomes more complex. In particular, depending on $h$ and easy axis direction, the transition of $\mathbf{m}$ to a new steady state may occur in such a way that the sign of the scalar product $\mathbf{m} \cdot \mathbf{e}_{a}$ is changed. Moreover, there can be a few changes of sign as the magnetic field frequency grows. In other words, increasing $\Omega$ can lead to repeated switching of the nanoparticle magnetization. Because each switching is accompanied by a strong change of $m_{z}$ (if $\theta_{a}$ is not too close to $\pi/2$), these switchings can appreciably affect the frequency dependence of $\langle \overline {m_{z}} \rangle_{ \mathrm{num}}$, see Fig.~\ref{fig5}. For illustration, in Fig.~\ref{fig6} we show examples of steady-state trajectories of $\mathbf{m}$ just before and after the switching transition. Note also that, if $h$ is large enough, there can exist a frequency interval, where the magnetization dynamics is chaotic, i.e., the time evolution of $\mathbf{m}$ is extremely sensitive to initial conditions. In our numerical calculations, the time-averaging interval $\mathcal {T}$ for nanoparticles with chaotic magnetization dynamics is chosen to be $2\cdot 10^{3} /\Omega$ and $4\cdot 10^{3} /\Omega$ for $\Omega<1$ and $\Omega \geq 1$, respectively. It turned out that frequency-induced transitions to and from chaotic regime do not change significantly the frequency dependence of $\langle \overline {m_{z}} \rangle_{ \mathrm{num}}$.

\section{POWER LOSS}
\label{Power}

According to the definition (\ref{def_q}) and the series representation (\ref{ser_m}), the reduced power loss at $h\ll 1$ and under the condition that the reduced magnetization $\mathbf{m}$ does not transit to another steady states with increasing $h$ and $\Omega$ can be expressed in the general form
\begin{equation}
    q = \alpha \langle \overline{\Big(\sum
    \nolimits_{n=1}^{\infty}\dot{\mathbf{m}
    }_{n}\Big)^{\!2}} \rangle.
    \label{q}
\end{equation}
For simplicity and illustrative purposes, we restrict ourselves to the second order in the expansion of $q$ in powers of $h$. In this approximation the above expression reads
\begin{equation}
    q = \alpha \langle \overline{\dot{\mathbf{m}}
    _{1}^{2}}\rangle.
    \label{q2_1}
\end{equation}
Since, according to (\ref{m1}), $\overline{\dot{\mathbf{m} }_{1}^{2}} = \Omega^{2} \overline{\mathbf{m}_{1}^{2}}$ and $\overline{\mathbf{m}_{1}^{2}}$ is given by (\ref{av_m1}), the reduced power loss can be written as $q = \alpha \Omega^{2} (\langle a_{11}^{2}\rangle + \langle b_{11}^{2}\rangle + \langle c_{11}^{2}\rangle + \langle d_{11}^{2}\rangle)$. Calculating these averages, which can be done by using expressions (\ref{abcd_11}) together with the conditions $\langle \kappa_{a}^{2} \rangle = \langle \delta_{a}^{2} \rangle = 1/2$ and $\langle \lambda_{a}^{2} \kappa_{a}^{2} \rangle = \langle \lambda_{a}^{2} \delta_{a}^{2} \rangle = 1/6$, we obtain
\begin{equation}
    q = \frac{1}{3}\alpha (1+\rho^{2})h^{2}
    \frac{\Omega^{2} (1 + \ell \Omega^{2})}
    {(1 - \ell \Omega^{2})^{2} + 4\alpha^{2}
    \Omega^{2}}.
    \label{q2_2}
\end{equation}

\begin{figure}
    \centering
    \includegraphics[totalheight=5cm]{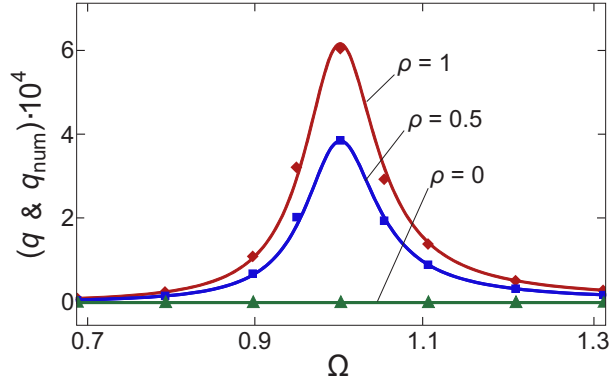}
    \caption{\label{fig7} (Color online) Frequency
    dependence of the reduced power loss for the
    circularly ($\rho = 1$), elliptically ($\rho =
    0.5$), and linearly ($\rho=0$) polarized magnetic
    field of amplitude $h=0.01$. The numerical results
    ($q_{\mathrm{num}}$) obtained by solving
    Eq.~(\ref{red_LLG}) are represented by symbols,
    and the solid lines represent the theoretical
    result (\ref{q2_2}).}
\end{figure}

\begin{figure}
    \centering
    \includegraphics[totalheight=5cm]{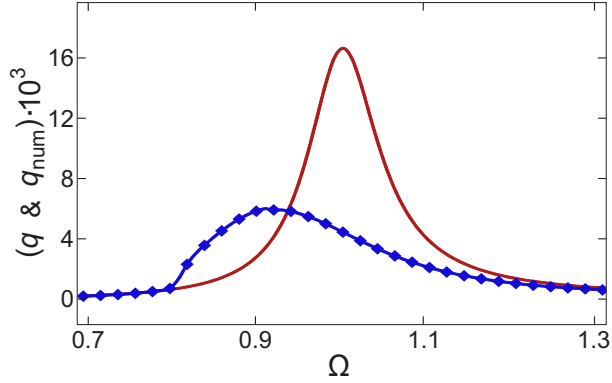}
    \caption{\label{fig8} (Color online) Frequency
    dependence of the reduced power loss for $h =
    0.05$ and $\rho = 1$. The theoretical ($q$) and
    numerical ($q_{\mathrm{num}}$) results are shown
    by the solid and symbol lines, respectively. The
    difference between $q$ and $q_{\mathrm{num}}$
    arises from the same transitions that are
    responsible for the difference between $\langle
    \overline{m_{z}} \rangle$ and $\langle \overline
    {m_{z}}\rangle_{\mathrm{num}}$, see Fig.~\ref{fig3}.}
\end{figure}

\begin{figure}
    \centering
    \includegraphics[totalheight=5cm]{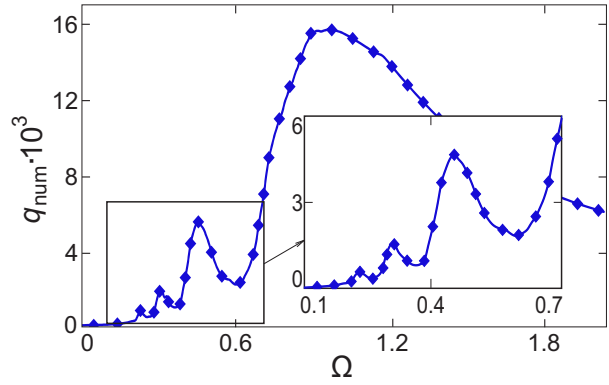}
    \caption{\label{fig9} (Color online) Dependence
    of $q_{\mathrm{num}}$ on the reduced frequency
    $\Omega$ for $h = 0.3$ and $\rho = 1$. Insert:
    the same dependence in the vicinity of the
    fourth-order ($\Omega =1/4$), third-order
    ($\Omega =1/3$), and second-order ($\Omega =
    1/2$) resonances.  The shift of the peak positions
    of $q_{\mathrm{num}}$ to the left has the same
    origin as the shift of the minima of $\langle
    \overline  {m_{z}} \rangle_{  \mathrm{num}}$.}
\end{figure}

Thus, in the second-order approximation, the reduced power loss $q$ as a function of the reduced frequency $\Omega$ exhibits a resonant behavior near the point $\Omega =1$ and satisfies the conditions $\max{q} \simeq (1 + \rho^{2}) h^{2}/(6 \alpha)$ at $\alpha \ll 1$ and $q \simeq \alpha (1 + \rho^{2}) h^{2}/3$ at $\Omega \gg 1$ and $\alpha \ll 1$. This is not surprising because the second-order expansion of the power loss corresponds to the first-order expansion of the magnetization. To check formula (\ref{q2_2}), we numerically calculated the reduced power loss as $q_{\mathrm{num}} = (1/N \mathcal {T}) \sum_{l=1}^{N} \int_{0}^{\mathcal {T}} q^{(l)} d\tau$, where $q^{(l)} = \alpha (\dot{\mathbf{m} }^{(l)})^{2}$. If the reduced magnetic field amplitude $h$ is small enough, the numerical results are in excellent agreement with the analytical ones, as seen from Fig.~\ref{fig7}. For the same reasons as for $\langle \overline {m_{z}} \rangle$, the increase of $h$ leads to the difference between $q$ and $q_{\mathrm {num}}$, see Fig.~\ref{fig8} for an illustration.
If the magnetic field amplitude is not too small, the nonlinear resonances can modify the frequency dependence of the power loss. In particular, assuming that $\dot{\mathbf{m}} = \sum_{n=1}^{3} \dot{\mathbf{m}}_{n}$ and taking into account that according to (\ref{m1}) and (\ref{m2}) $\overline{\dot{ \mathbf{m} }_{1}\cdot \dot{ \mathbf{m}}_{2}} = 0$, one gets (up to terms of order $h^{4}$)
\begin{equation}
    q = \alpha \langle \overline{\dot{\mathbf{m}}
    _{1}^{2}} + \overline{\dot{\mathbf{m}}_{2}^{2}}
    + 2\overline{\dot{\mathbf{m}}_{1}\cdot
    \dot{\mathbf{m}}_{3}} \rangle.
    \label{q_four}
\end{equation}
As is clear from the above discussion, the terms in $\dot{\mathbf{m}}_{1}\cdot \dot{\mathbf{m} }_{3}$ that show a resonant behavior in the vicinity of the reduced frequency $\Omega = 1/3$ vanish upon time averaging. Therefore, the only term $\alpha \langle \overline{\dot{\mathbf{m}}_{2}^{2}} \rangle$, which is of the order of $h^{4}$, may qualitatively change the frequency dependence of the reduced power loss (\ref{q2_2}), which is of the order of $h^{2}$. According to (\ref{abcd_21})--(\ref{m2}), this occurs near the second-order resonance, i.e., in a small vicinity of the reduced frequency $\Omega = 1/2$. A similar analysis predicts and numerical results confirm, see Fig.~\ref{fig9}, that the higher-order resonances can also exist. Note that these resonances are more pronounced for $q_{\mathrm{num}}$ than for $\langle \overline {m_{z}} \rangle_{ \mathrm{num}}$.

\section{DISCUSSION AND CONCLUSIONS}
\label{Concl}

We have determined the average magnetization and power loss for the system of ferromagnetic nanoparticles that are driven by an elliptically polarized magnetic field and whose anisotropy axes are uniformly oriented. One of the most important observations is that the driving field magnetizes this system in the direction perpendicular to the polarization plane. This is a remarkable result because the elliptically polarized magnetic field has no component in that direction. From a physical point of view, the appearance of the average magnetization is a consequence of the fact that the magnetization precession in nanoparticles occurs in the counterclockwise direction. Indeed, due to this property, the magnetization dynamics in each pair of nanoparticles, whose equilibrium magnetization vectors are symmetric with respect to reflection in the polarization plane, is different. This difference is a purely dynamical, polarization-dependent effect, which after averaging over all nanoparticles leads to a non-zero average magnetization of the reference system.

In order to find the analytical expressions for the average magnetization and power loss in the case of small-amplitude limit of the driving magnetic field, we have developed a general perturbation theory for the Landau-Lifshitz-Gilbert (LLG) equation. Within this framework, we have determined the steady-state solution of the LLG equation and calculated the average magnetization and power loss with the second-order accuracy. An important feature of these quantities is that they depend on the driving field frequency in a resonant way. It should be emphasized that, according to the definition, the second-order expression for the power loss follows from the first-order solution of the LLG equation, and so exhibits a resonant behavior in the vicinity of the first-order resonance. In contrast, the second-order expression for the average magnetization is determined by the second-order solution of the LLG equation. Although this solution accounts for the effect of both the first- and second-order resonances, the impact of the second-order resonance is eliminated by the averaging. We have confirmed these theoretical predictions by the numerical results obtained from numerical solution of the LLG equation.

Our theoretical analysis has shown, and numerical results have verified, that subharmonic resonances arising from the nonlinearity of the LLG equation also influence the frequency dependence of the average magnetization and power loss. However, since subharmonic resonances appear for rather large amplitudes of the elliptically polarized magnetic field, the nonlinear features of the magnetization dynamics strongly influence the frequency dependence of the reference quantities as well. We have found that among these features the transitions between different steady-state solutions of the LLG equation, which occur as the driving field frequency changes, play the most important role. If these transitions occur without the magnetization switching, the extremes of the average magnetization and power loss are shifted to lower frequencies. In contrast, if the transitions in some fraction of nanoparticles are accompanied by the magnetization switching (this is possible if the driving field amplitude is large enough), then the frequency dependence of these quantities, and above all the average magnetization, changes drastically. Finally, we have established that the transitions between regular and chaotic regimes of the magnetization dynamics do not affect these quantities in a significant way.

Let us also discuss the nanoparticle systems that can be used to verify the obtained results. According to the model assumptions, the experimental systems must be composed of monodisperse single-domain nanoparticles that are randomly oriented and do not interact with each other. While the systems with monodisperse single-domain nanoparticles are common and easy to synthesize,\cite{FPCS} the systems characterized by the uniform distribution of easy axis directions and negligible dipolar interaction are not so widespread. To the best of our knowledge, one of the most suitable systems, whose magnetic dynamics can be described by the proposed model, is the assembly of iron-platinum nanoparticles produced at relatively low annealing temperature.\cite{Sun} Another such system is the two-dimensional assembly of iron oxide nanoparticles obtained by the click reaction.\cite{TPCM} If the interparticle distance is large enough, the nanoparticles in this assembly satisfy all of the above conditions.

\section*{ACKNOWLEDGMENTS}

The authors are grateful to the Ministry of Education and Science of Ukraine for financial support under Grant No.\ 0116U002622.

\end{document}